\documentclass[%
 reprint,
superscriptaddress,
 amsmath,amssymb,
 aps,
floatfix,
showkeys
]{revtex4-2}

\usepackage{physics}
\usepackage{graphicx}% Include figure files
\usepackage[caption=false]{subfig}
\usepackage{dcolumn}% Align table columns on decimal point
\usepackage{bm}% bold math
\usepackage{hyperref}% add hypertext capabilities

\usepackage{tikz}
  \usetikzlibrary{positioning}
  \usetikzlibrary{quantikz}

\usepackage{amsmath}
\usepackage{mathtools}

\DeclareMathOperator*{\argmin}{arg\,min}

\tikzset{
operator/.append style={rounded corners},
}

\begin{document}

\title{Quantum neural network autoencoder and classifier applied to an industrial case study}

\author{Stefano Mangini}
\email{stefano.mangini01@universitadipavia.it}
\affiliation{Dipartimento di Fisica, Università di Pavia, Via Bassi 6, I-27100, Pavia, Italy}
\affiliation{INFN Sezione di Pavia, Via Bassi 6, I-27100, Pavia, Italy}

\author{Alessia Marruzzo}
\affiliation{Eni SpA, via Emilia 1, I-20097, San Donato Milanese, Italy}

\author{Marco Piantanida}
\affiliation{Eni SpA, via Emilia 1, I-20097, San Donato Milanese, Italy}

\author{Dario Gerace}
\affiliation{Dipartimento di Fisica, Università di Pavia, Via Bassi 6, I-27100, Pavia, Italy}

\author{Daniele Bajoni}
\affiliation{Dipartimento di Ingegneria Industriale e dell’Informazione, Università di Pavia, Via Ferrata 1, 27100, Pavia, Italy}

\author{Chiara Macchiavello}
\affiliation{Dipartimento di Fisica, Università di Pavia, Via Bassi 6, I-27100, Pavia, Italy}
\affiliation{INFN Sezione di Pavia, Via Bassi 6, I-27100, Pavia, Italy}
\affiliation{CNR-INO -  Largo E. Fermi 6, I-50125, Firenze, Italy} 

\date{\today}


\begin{abstract}
\section*{Abstract}
Quantum computing technologies are in the process of moving from academic research to real industrial applications, with the first hints of quantum advantage demonstrated in recent months.
In these early practical uses of quantum computers it is relevant to develop algorithms that are useful for actual industrial processes. In this work we propose a quantum pipeline, comprising a quantum autoencoder followed by a quantum classifier, which are used to first compress and then label classical data coming from a separator, i.e., a machine used in one of Eni's Oil Treatment Plants. This work represents one of the first attempts to integrate quantum computing procedures in a real-case scenario of an industrial pipeline, in particular using actual data coming from physical machines, rather than pedagogical data from benchmark datasets.  
\end{abstract}

\maketitle

\section{\label{sec:intro} Introduction}
We are currently witnessing a time of intense growth and investments into quantum computing technologies, both from academic and private sectors, aimed at a fast pace of advancement in the quest for computational advantage brought by the practical use of  quantum information processing.
In particular, it is believed we are now experiencing what has been termed the Noisy Intermediate Scale Quantum (NISQ) Computing era, i.e., quantum processing units (QPU) are available with a number of non-error corrected qubits scaling between 50 and 1000 \cite{Preskill2018NISQ}. While not allowing to perform fault tolerant quantum computing, these devices are becoming worldwide available to explore the frontiers of quantum algorithms, which exploit inherent quantum mechanical features such as superposition and entanglement to produce a radically different approach to computational problems \cite{Nielsen_Chuang}.

There are several fields of application in which new quantum algorithms have been analyzed: quantum chemistry~\cite{SokolovQuantumChemistry, BarkoutsosQuantumChemistry2018, PeruzzoVQE2014, Kivlichan2018QuantumChemistry}, optimization problems~\cite{farhi2014quantum, QAOA_Venturelli}, machine learning\cite{Biamonte2017QML, Havlicek2019SVM, Abbas2020Power}, solution of linear problems~\cite{Harrow2009LinearSystem, bravoprieto2020VariationalLinearAlgebra, xu2019VariationalLinearAlgebra} and differential equations~\cite{ChildsQuantumDifferential2020}. To overcome the problems due to the limited number of qubits available and to the absence of efficient error correction techniques, several proof-of-principle demonstrations have been carried out by focusing on so called variational quantum algorithms, characterized by a hybrid approach in which the quantum processing units (QPU) is seen as an accelerator alongside the classical CPU \cite{McClean2016VQAs, Cerezo2020VQAReview, Bharti2021NISQ}. 
Several of these studies also fall within the emerging field of Quantum Machine Learning \cite{Biamonte2017QML, Benedetti2019Parametrized, ManginiPerspective2021}, which has even triggered the birth of dedicated quantum machine learning software \cite{Pennylane, Broughton2020TensorflowQuantum, Qiskit}. 
%and new software are already available (Zapata, 2019) (Xanadu, 2019). 
Currently, most quantum machine learning algorithms are based on parametrized quantum circuits, and leverage an approach in which the optimization over the variational parameters is done on the classical CPU \cite{Benedetti2019Parametrized, ManginiPerspective2021}. Indeed, these parametrized quantum circuits, often referred to as \textit{quantum neural networks}~\cite{Schuld2014Quest, ManginiPerspective2021, McClean2018Barren} prove robust even in the presence of noise~\cite{McClean2016VQAs, Sharma_2020, Gentini_Noise}, which is inevitable in current implementation of quantum hardware, and are thus well-suited for near term NISQ devices. 

Here we test the use of quantum machine learning algorithms on a specific industrial use case. In particular, we propose the application of a newly formulated quantum pipeline comprising a quantum autoencoder algorithm~\cite{RomeroQAE2017, Bravo_2021quantum, Lamata_2018, Khoshaman_2018} followed by a quantum classifier, applied to real data coming from a first stage water/oil separator of one of Eni’s oil treatment plant. This algorithm is compared to the performance of a classical autoencoder to compress the original data, which are then used to implement a classification task. It is particularly relevant to notice that these quantum autoencoding algorithms can be run on presently existing quantum hardware, thus making such quantum machine learning algorithm readily usable with actual input data coming from a realistic source of industrial interest. While various models of variational autoencoders in the quantum domain have been proposed in the literature, for example for generative modelling tasks~\cite{Khoshaman_2018} and for the study of entanglement in quantum states~\cite{Chen_2021}, our implementation of the quantum autoencoder directly follows the architecture proposed by authors in~\cite{RomeroQAE2017}, which is often studied as a prototypical model in the quantum machine learning literature~\cite{Cerezo2020CostBarren}, and it was also even extended to feature input redundancy~\cite{Perez2020Reuploading}, as discussed in~\cite{Bravo_2021quantum}.

The manuscript is organized as follows. In Sec.~\ref{sec:case_study} we explain and give the specifics of the industrial case study considered in this work. In Sec.~\ref{sec:autoencoder} we introduce the classical neural network model of the autoencoder, and also discuss the clustering algorithm used to create the two classes for the classification problem. 
In Sec.~\ref{sec:quantumautoencoder} we review the quantum algorithm developed for a continuously valued input neuron \cite{ManginiQNeuronl2020}, from which the quantum algorithm for the quantum autoencoder is derived. In Sec.~\ref{sec:results} we show the results obtained for the data compression task, comparing them with those obtained with the purely classical autoencoder. At last in Sec.~\ref{sec:classification}, we use the compressed data to implement a quantum classifier used to label the original data in a binary classification problem.

\section{\label{sec:case_study} Case study}
\begin{figure}
    \centering
    \includegraphics[width=\columnwidth]{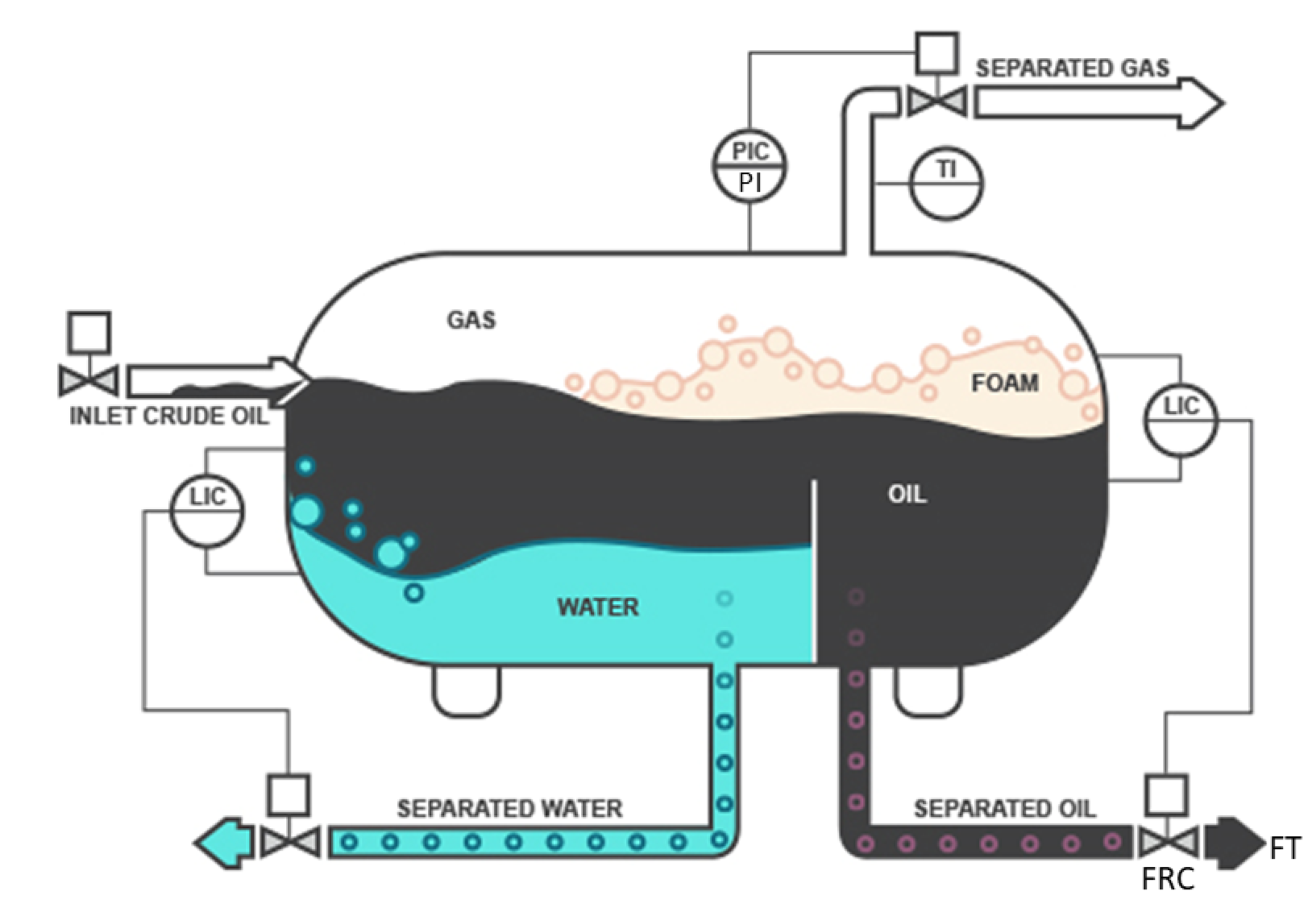}
    \caption{Snapshot of the separator. The separator is regulated with three controllers: a pressure controller for the output gas stream, and two-level controllers for the water and the oil stream. The controllers use PID controller equations to regulate the opening of valves on the output streams.}
    \label{fig:separator}
\end{figure}

The industrial case study discussed in this work aims at testing classical and quantum machine learning approaches to analyze data coming from an industrial equipment within one of Eni's Oil Treatment plants, showed in Fig.~\ref{fig:separator}. 

The equipment is a separator, i.e. a vessel receiving a stream of high pressure, high temperature crude oil (left part of the figure, indicated with a black stream), and exploits gravity to separate three output streams: Water (the heaviest component), indicated in the figure with a light blue stream; Oil (intermediate component), in the lower part of the figure indicated with a black stream; and Gas (lightest component), indicated with a light grey stream. The separator is regulated with three controllers: a pressure controller for the output gas stream, and two-level controllers for the water and the oil stream. Notice that the controllers use PID (proportional – integral – derivative) controller equations to regulate the opening of valves on the output streams. 

In a realistic machine learning problem, we might wish to use all the measurements coming from the sensors installed on this component, as well as on some of the components installed upstream, in order to predict if the behavior of the equipment is normal or faulty (i.e. working in a degraded mode). However, due to the limitation in the complexity of the problems that can currently be faced with quantum computing, we will focus on a simplified problem, involving only 4 variables, that are: 
\begin{itemize}
\item the oil level (LIC),
\item the oil output flow (FT),
\item the pressure (PI),
\item the opening of the oil output valve (FRC).
\end{itemize}

Sensor measurements are sampled every $10$ seconds and stored into data tables to be used for the training of the neural networks. 

The first step of the case study is the implementation of a dimensionality reduction procedure to compress the 4-dimensional input vector $\vec{x}=(x_\text{FRC},\, x_\text{FT},\, x_\text{LIC},\, x_\text{PI})$ into a $2$-dimensional vector. This is done both via a standard classical neural network autoencoder and a quantum autoencoder, introduced in Sec.~\ref{sec:autoencoder} and Sec.~\ref{sec:quantumautoencoder} respectively. 

\begin{figure*}[htbp]
   \includegraphics[width=\textwidth]{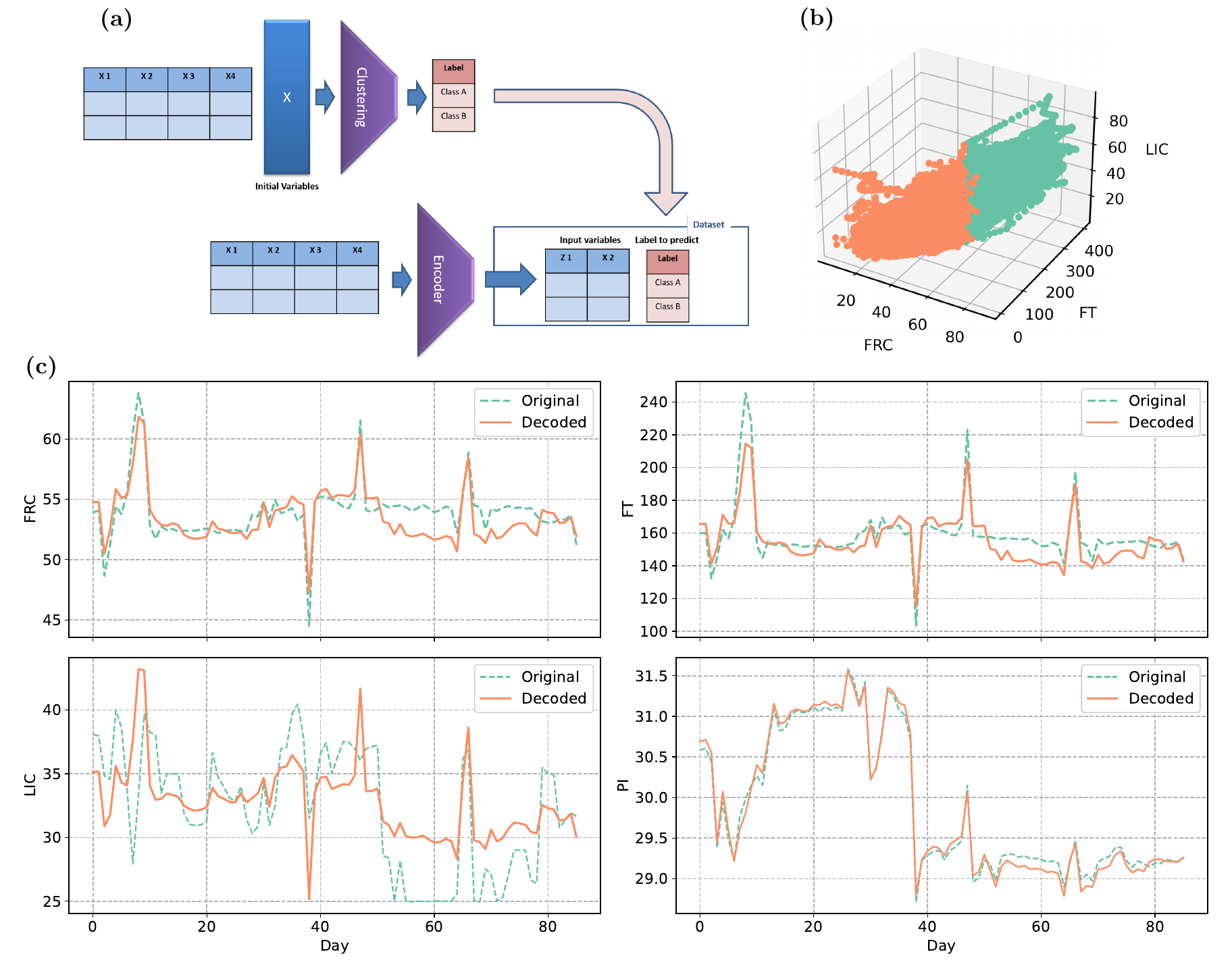}
    \caption{\textbf{(a)} The approach followed in this project: a clustering algorithm was used to define two categorical classes (Class A and Class B). Then, an autoencoder was used to reduce the dimensionality of the problem. Finally, a classifier was used to predict Class A and Class B identified with the clustering algorithm. \textbf{(b)} Results of the clustering algorithm KMeans on the input data. In particular, only the features FRC, FT and LIC are shown. The different color indicates the different label (or class) assigned to the data. \textbf{(c)} Plot of the decoded data on top of the original validation data averaged by day. Here the features were rescaled to their original range.}
    \label{fig:FIG3}
\end{figure*}

The second step will be to implement a classifier using the $2$-dimensional latent vector from the compression step to classify the status of the component. In order to do so, we need a labeled training dataset associating an input $\vec{x}_i$ to a label $y_i=\{0,1\}$ corresponding to the ``ok" or ``faulty" state respectively. However, since $4$ variables are too few to label the working status of the separator as ``ok" or ``faulty", we followed a different approach, as explained in the upper left panel of Fig.~\ref{fig:FIG3}. We run a binary clustering algorithm on the initial variables, in order to identify two categorical states, named as ``Class A” and ``Class B”, and then used these categorical states as the labels for the classification task. So, the latent vector from the encoder is used as input for the classifier, that is trained to correctly predict the ``Class A” and ``Class B” states. The clustering algorithm used is the KMeans algorithm as implemented in the \texttt{scikit-learn} library~\cite{scikit-learn}. This algorithm takes as input the desired number of clusters, in our case two, and tries to split the data in groups of equal variance. The centroids of the clusters were initialized uniformly at random. In Fig.~\ref{fig:FIG3} we show the result of the clustering procedure, where for ease of plotting we show only three of the four variables. This categorical dataset is then used to train a classical and quantum classifier, whose implementation details and results are discussed in Sec.~\ref{sec:classification}.

In Table~\ref{tab:ex} we summarize the findings of our work, showing the key figures (compression error and classification accuracy) for the classical and quantum pipelines considered in the case study. 

\begin{table}[htbp]
\caption{\label{tab:ex} Key figures for the compression and classification tasks for the classical and quantum procedures considered. The compression task is implemented with classical and quantum autoencoders; the classification task is implemented with a KNeighborsClassifier and with a single qubit variational classifier.}
\begin{ruledtabular}
\begin{tabular}{ccc}
          & Compression & Classification \\
          & [error]\footnote{Average reconstruction error, as defined in Eq.~\eqref{eq:recontruction_error}.} &  [accuracy]\footnote{Classification accuracy, defined as the percentage of correctly classified data.} \\
Classical & 5\%         & 89.7\% \\
Quantum   & 5.4\%       & 87.4\% \\
Quantum hardware (\texttt{ibmq\_x2}) & --- & 82.3\% \\
\end{tabular}
\end{ruledtabular}
\end{table}

\section{\label{sec:autoencoder} Neural network autoencoder}
\begin{figure}[htbp]
    \centering
   \includegraphics[width=\columnwidth]{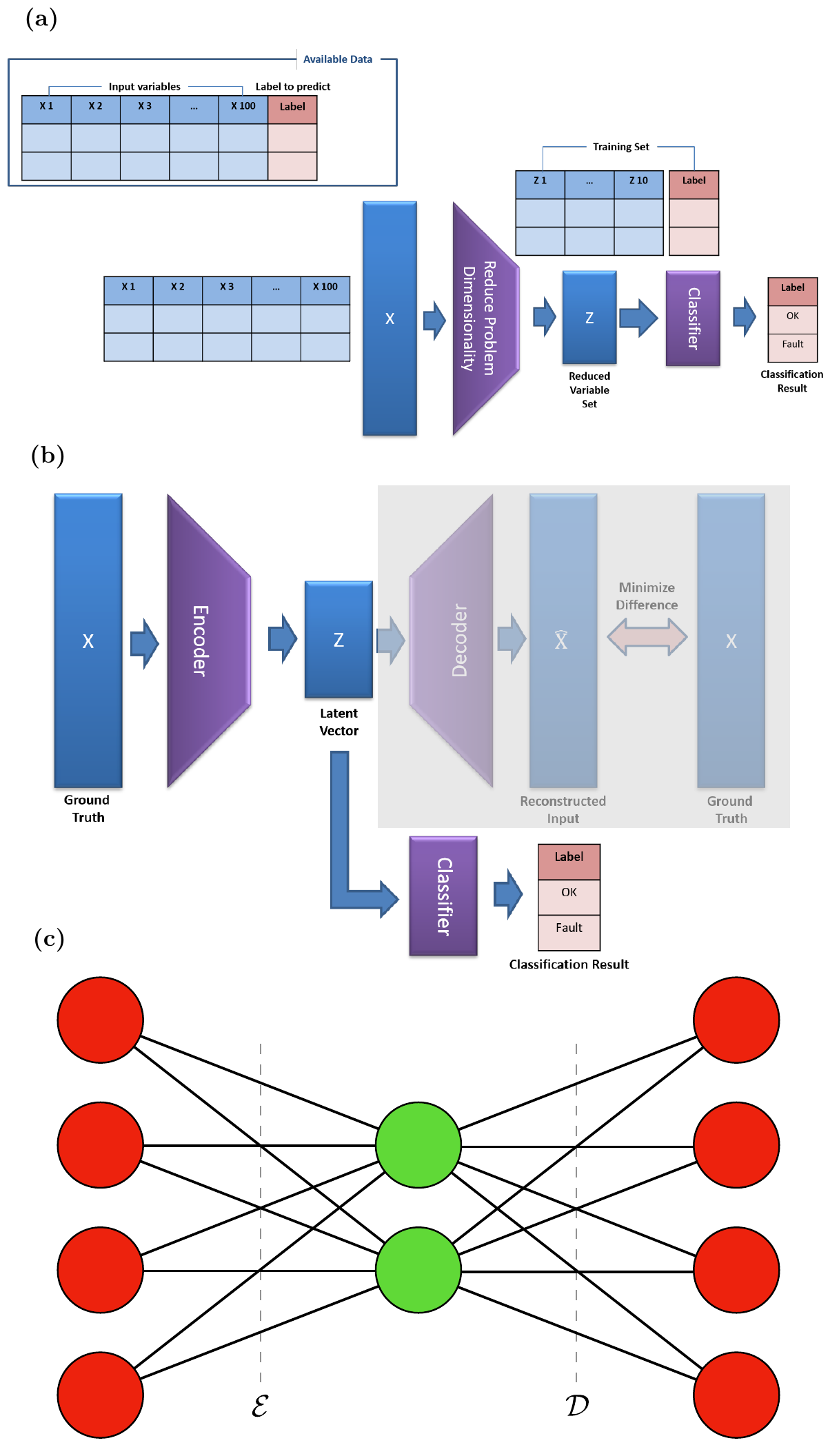}
    
    \caption{\textbf{(a)} Reducing the dimensionality of a classification problem. \textbf{(b)} Using Autoencoders to reduce the dimensionality of a problem and solving the classification problem on the reduced variable set. \textbf{(c)} Schematic representation of the neural network autoencoder architecture. The input neurons in red are mapped to an hidden layer (in green) of lower dimension, storing the compressed information. Then, an output layer with the same number of neurons as the input one, tries to restores the original data with low error.}
    \label{fig:FIG1}
\end{figure}

The most common use case of artificial neural networks is supervised learning, where the network is asked to learn a mapping from an input to an output space, by having access to an example set of input-output pairs. One prominent example are classification tasks, where the network is presented a labeled dataset $\mathcal{M} = ((\vec{x}_1, y_1), (\vec{x}_2, y_2), \hdots, (\vec{x}_M, y_M)) \subset \mathbb{R}^n \times \{0,1\hdots,c\}$ consisting of a set of inputs $\vec{x}_i$ and the corresponding correct labels $y_j$, with $c$ being the total number of classes the inputs can be divided into (see upper left side of Fig.~\ref{fig:FIG1}). Using this dataset, called \textit{training set}, a neural network can be trained in a supervised fashion to learn the relationship between the input variables and the expected classification results. When the training is complete, the neural network model can be used for \textit{inference}, that is for labeling previously unseen data. This property of neural networks, called \textit{generalization}, is ultimately the key figure that distinguishes them from standard fitting techniques, making them incredibly powerful tools\cite{Goodfellow2016DeepL, LeCun2015DeepL, Hastie2009Statistical, HumanlevelControlDeep2015a}.  

When dealing with real world problems, such as classifying the operational status of a plant as ``ok" or ``faulty" based on the measurements from the sensors installed on the plant, it is often the case  that a large number of input variables are available. In fact, measurements coming from tens of sensors need to be analyzed not only on their instantaneous values, but also on additional features computed on time intervals, such as moving averages, and minimal/maximal values trends. This leads to a situation where too many input variables are available in the dataset, and it is often ineffective to directly feed them into the neural network classifier. With such a large number of variables, correlation analysis and feature engineering are often performed to focus only on the most influencing variables, and only after these pre-processing steps the neural network can be used effectively. Another strategy is to use a \textit{dimensionality reduction} approach, consisting in computing a new set of variables, smaller than the initial one, incorporating most –ideally all– of the information contained in the original data. These new compressed data are then used as inputs to the classifier, as shown in  Fig.~\ref{fig:FIG1}a. 

In order to reduce the problem dimensionality, methods such as PCA (Principal Component Analysis) or SVD (Singular Value Decomposition)~\cite{Hastie2009Statistical} are typically used. However, these methods are based on linear decomposition of the initial variable space, and they could not be suitable when nonlinear relationships between the variables need to be kept into account.

\subsection{Classical Autoencoders}

An alternative method to reduce the dimensionality of the problem is to use  Autoencoders~\cite{Goodfellow2016DeepL}, as shown in Fig~\ref{fig:FIG1}(c). An autoencoder is a neural network composed of two modules, called \textit{encoder} and \textit{decoder}, designed in such a way that the subsequent application of the encoder and the decoder to the input data results into an output that is as close as possible to the input, i.e. the discrepancy between output and input is minimized. With such an approach, the encoder builds a compressed representation of the input data to be eventually used by the decoder to fully (and as faithfully as possible) reconstruct the input. This means that the compressed representation built by the encoder (often referred to as latent vector) contains the same information of the initial input space, or at least minimum information is lost.

Once the autoencoder has been trained to reconstruct the input, the latent vector can be used as the input space for the classifier. Therefore, the classification problem can be described as shown in  Fig.~\ref{fig:FIG1}b. 

In our case study, we consider a neural autoencoder as shown in  Fig.~\ref{fig:FIG1}c. The original input variables are fed to the input neurons, which are then passed to an intermediate hidden level (shown in green) consisting of a number of neurons much smaller than the input. Finally, there is an output layer (shown in red) with the same number of neurons as the input. The neural network is trained in an unsupervised fashion in order to generate an output that is as close as possible to the input. Thus, if it is possible to reconstruct the input (with a minimum loss of fidelity) starting from the inner layer, this means that the inner layer contains the same information as the input, and therefore we can use the compressed layer as an input for the classifier. The presence of non-linear activation functions within the neural network, such as the Rectified Linear Unit $\text{ReLU}(x) = \max(0, x)$, or sigmoid $s(x) = 1/(1+e^{-x})$, ensures that the network can better capture non-linear relationships in the input variables compared to PCA or SVD.

\section{\label{sec:quantumautoencoder} Quantum Data Compression}
In order to use a quantum pipeline to analyze the classical data coming from the sensors, we need to encode such data on a quantum state to be used as the input of the quantum autoencoder. While it is known from the recent literature \cite{Abbas2020Power, Lloyd2020Embeddings, Schuld2020Encoding, Theis2020Expressivity, LaRose2020Robust, Mitarai2018Learning} that choosing a good encoding scheme is of key importance to ensure good expressivity and representation power of variational quantum algorithms, there is still no standard procedure to do so. In our case, given the relatively simple and low dimensional nature of the data sets to be analyzed, we choose to use a phase encoding strategy \cite{TacchinoQNeuron2019,ManginiQNeuronl2020}, which provides an effective way to load classical data into a quantum state, and also already proved useful in other machine learning tasks such as pattern-recognition~\cite{TacchinoQNeuron2019,ManginiQNeuronl2020,Tacchino2020Network,Tacchino2020IEEE}. In particular, given a data sample $\vec{x} = (x_1, x_2, \hdots, x_N) \in \mathbb{R}^N$, this is encoded on the quantum state of $n=\log_2 N$ qubits as follows 
\begin{equation}
\ket{\psi_{\vec{x}}} = \sum_{i=1}^{2^n}e^{i\, x_i}\ket{i}
\label{eq:phase_encoding}
\end{equation}
where the data $\vec{x}$ are first re-scaled to fit into an appropriate range, such as $ x_i \in [0, \pi]$. This class of states is also known as \textit{locally maximally entangled} (LME) states \cite{KrausLME2009}, and we refer to Refs.~\cite{ManginiQNeuronl2020, TacchinoQNeuron2019} for an extended discussion on these states for variational quantum procedures.

\subsection{\label{ssec:qae} Quantum Autoencoder}
Having fixed a data encoding strategy, we now build a variational quantum algorithm for data compression. In particular, borrowing from the classical machine learning literature, our goal is to implement a quantum autoencoder \cite{RomeroQAE2017,Lamata_2018,Bravo_2021quantum}. In classical autoencoders, the compression is built in the geometric structure of the neural network, since the input layer is followed by a much smaller hidden layer consisting of a number of neurons equal to the desired reduced dimension. This bottleneck forces the NN to learn a low dimensional representation of the inputs, which is stored in the intermediate hidden layer(s) of the network. However, this procedure cannot be straightforwardly applied to the quantum domain, because quantum computations follow a unitary, thus reversible, evolution. In fact, while classically it is possible to perform {\texttt{fan-in}}({\texttt{fan-out}}) operations, that is arbitrarily reducing (increasing) the number of classical bits in the computation, such operations are irreversible, which prevents their direct implementation on a quantum computer. Alternatively said, it is not possible to eliminate or create new qubits during the execution of a quantum computation. 

Nonetheless, it is possible to circumvent this issue as follows. Consider two quantum systems, denoted as system $A$ and system $B$, and be $\ket{\psi}_{AB}$ the quantum state of the composite quantum system $AB$. Our goal is to compress the information stored in the composite state in a lower dimensional representation, for example given by the state of subsystem $A$ only, with system $B$ being safely discarded. We can formalize this intuition in the following way: denote with $\mathcal{E}(\bm{\theta})$ a quantum encoding (in the sense of \textit{compressing}) operation depending on variational parameters (i.e. trainable weights) $\bm{\theta}$, then the desired compression task consists in the operation
\begin{equation}
\mathcal{E}(\bm{\theta})\ket{\psi}_{AB}=\ket{\phi}_A \otimes \ket{\rm{trash}}_B \, ,
\label{eq:q_encoder}
\end{equation}
where the state $\ket{\psi}_{AB}$ of the composite system $AB$ is compressed on the state $\ket{\phi}_A$ of subsystem A only, and the system $B$ is mapped to a fixed reference state of choice, called \textit{trash} state, for example being the ground state $\ket{\rm{trash}}_B = \ket{0}^{\otimes |B|}$. It is clear that the goal of the encoder is  to \textit{disentangle} the two systems in such a way that one of them, i.e., the trash system, goes to the fixed reference state, and the other  contains all the original information of from the full quantum state. In order to recover the original quantum state $\ket{\psi}_{AB}$, it is then possible to act with a \textit{quantum decoder} operation $\mathcal{D}(\bm{\theta})$, defined as $\mathcal{D}(\bm{\theta})=\mathcal{E}^{\dagger}(\bm{\theta})$. This way, acting with the decoder on the compressed state yields the original state
\begin{eqnarray}
       \mathcal{D}(\bm{\theta})(\ket{\phi}_A \otimes \ket{\rm{trash}}_B) &=& \mathcal{D}(\bm{\theta})\, \big(\mathcal{E}(\bm{\theta}) \ket{\psi}_{AB}\big) \nonumber\\
    &=& \big(\mathcal{E}(\bm{\theta})^\dagger \mathcal{E}(\bm{\theta})\big)\, \ket{\psi}_{AB} \nonumber\\
    &=& \ket{\psi}_{AB}\, . \nonumber
\end{eqnarray}

\begin{figure}
    \centering
   \includegraphics[width=\columnwidth]{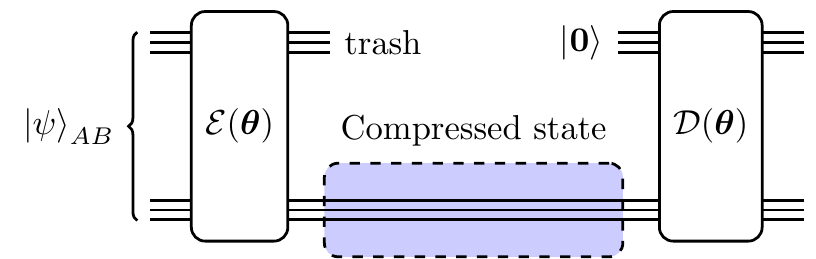}

    \caption{Schematic representation of the generic quantum autoencoder algorithm. The input quantum state, $\ket{\psi}_{AB}$, is disentangled and the state of the system $B$,  defined ``trash'' system, is mapped to a reference quantum state, $\ket{\bm{0}}_B$. The action may then be reversed by applying a quantum decoder operation $\mathcal{D}(\bm{\theta}) = \mathcal{E}^\dagger(\bm{\theta})$.}
    \label{fig:quantum_autoencoder}
\end{figure}

Thus, suppose having compressed the information stored in the quantum state of a composite system into one of its subsystems. Then, it is always possible to retrieve the original information, if needed, by coupling such information-carrying system with some new qubits initialized in the $\ket{\rm{trash}}$ state, and then act on them with the quantum decoder operator, as schematically represented in Fig. \ref{fig:quantum_autoencoder}. 

Of course, this only holds in the ideal case where the encoder perfectly manages to disentangle the subsystems $A$ and $B$, i.e., by obtaining the product state in Eq.\eqref{eq:q_encoder}. In practice, this is never the case since the input state $\ket{\psi}_{AB}$ depends on the classical input data via the phase encoding, and these states cannot be exactly disentangled, in general. In fact, after discarding the trash system $B$, the compressed state $A$ is no more a pure state, rather a mixed state given by the density matrix $\rho_A = \Tr_B [(\mathcal{E}(\bm{\theta})\ket{\psi_{AB}}) (\bra{\psi_{AB}})\mathcal{E}(\bm{\theta})^\dagger)]$. However, upon optimization of the variational parameters $\bm{\theta}$, the trained encoder tries to create a final state as close as possible to the target product state in Eq.~\eqref{eq:q_encoder}. 

\paragraph*{Training the quantum autoencoder}

The initial quantum state $\ket{\psi}_{AB}$ is obtained by using phase encoding to load the classical information on the phase of the quantum state, with the following scheme. Be $\mathcal{X} = \{\vec{x}_i\,|\, \vec{x}_i \in \mathbb{R}^N, i = 1,\hdots, M\}$ the set containing the classical data to be analyzed, then the quantum autoencoder is trained using the quantum states obtained as $\mathcal{T} = \{\ket{\psi_{\vec{x}}} = \sum_i e^{i\,x_i}\ket{i}\,|\, \forall \vec{x} \in \mathcal{X}\}$. In our specific case, the classical data are four dimensional $N=4$ and thus we only need $n=\log_2 N = 2$ qubits to encode the data. This in turn implies that the compressed system $A$ and the trash subsystem $B$ consist of a single qubit each.
Given the input data, the variational parameters $\bm{\theta}$ of the encoder are optimized in order to rotate the trash qubit as close as possible to the target trash state, which we choose to be $\ket{\rm{trash}} = \ket{0}$. This is achieved by means of a training procedure whose aim is to find optimal parameters $\bm{\theta}^\ast$ such that the \textit{loss function} characterizing the task, $\mathcal{L}(\bm{\theta})$, is minimized. That is, the goal of training is to find 
\begin{equation}
\begin{split}
\bm{\theta}^\ast &= \argmin_{\bm{\theta}} \mathcal{L}(\bm{\theta})\\
\text{with} \quad  \mathcal{L}(\bm{\theta}) &= \frac{1}{M}\sum_{j=1}^{M}\big|1-\expval{Z_B}_j\big|\, ,
\end{split}
\label{eq:loss_fn}
\end{equation}
where we have defined
\begin{equation}
\expval{Z_B}_j = \mel{\psi_{\vec{x}_j}}{\mathcal{E}^\dagger(\bm{\theta})(\mathbb{I}_A \otimes Z_B)\mathcal{E}(\bm{\theta})}{\psi_{\vec{x}_j}}
\end{equation}
as the mean value of the Pauli operator $Z=\rm{diag}(1,-1)$ evaluated on the trash system $B$, after the encoder acted on the input quantum state, $\ket{\psi_{\vec{x}_j}}$. 

\begin{figure}
    \centering
     \includegraphics[width=\columnwidth]{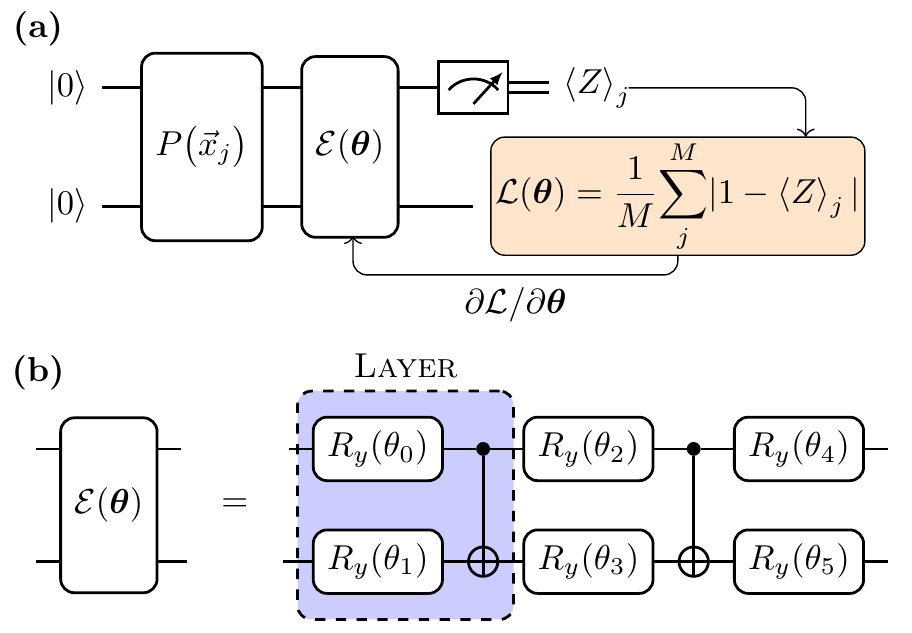}

    \caption{\textbf{(a)} Quantum circuit used to train the quantum autoencoder. A register initialized in the ground state $\ket{00}$ is first subject to the phase encoding operation denoted by $P(\vec{x})$, and then goes through the quantum encoder $\mathcal{E}(\bm{\theta})$. Then the trash qubit is measured, and the mean value of the Pauli operator $\expval{Z}$ is evaluated. Such value is then plugged into the loss function $\mathcal{L}(\bm{\theta})$ to drive the learning process. \textbf{(b)} Circuit representation of the quantum encoder $\mathcal{E}(\bm{\theta})$. Two layers of Pauli$-y$ rotations and CNOT, are followed by a final layer of Pauli$-y$ rotations. In total, the circuit has 6 trainable parameters. The decoder $\mathcal{D}(\bm{\theta})=\mathcal{E}^\dagger(\bm{\theta})$ is obtained by reversing the order of the operations, and changing the sign of the rotations angles.}
    \label{fig:qae_training_circuit}
\end{figure}

The loss function used in Eq.~\eqref{eq:loss_fn} is referred to as \textit{Mean Absolute Error} (MAE) in the classical machine learning literature, and together with the \textit{Mean Squared Error} (MSE) is the one of the most commonly employed loss functions in supervised regression tasks, which is also our case. Note that the loss function is faithful, in the sense that it reaches its global minimum $\mathcal{L}(\bm{\theta}^\ast) = 0$, only when $\expval{Z_B}_j=1,\,\forall j=1,\hdots,M$, that is when the trash qubit is always and perfectly disentangled from the other qubit, and mapped to the target trash state $\ket{0}$. A schematic representation of the quantum circuit used for the training procedure is explicitly shown in Fig.~\ref{fig:qae_training_circuit}(a). 

\paragraph*{Variational ansatz}
The actual quantum circuit implementation of the encoder $\mathcal{E}(\bm{\theta})$ (and hence the decoder) is arbitrary, and different variational ansatzes have been proposed in the quantum machine learning literature, in fact \cite{Cerezo2020VQAReview, ManginiPerspective2021, Bharti2021NISQ, McClean2016VQAs}. In our case, we are dealing with only two qubits, and the most general ansatz consists of repeated applications of single qubit rotations and CNOT quantum gates. In fact, having in mind to keep the parameters count and the overall circuit complexity low, we hereby propose a minimal yet efficient variational autoencoder consisting of two layers of Pauli$-y$ rotations $R_y(\theta)=e^{i\sigma_y\theta/2}$ and a CNOT, followed by a final layer of rotations, as schematically depicted in Fig.~\ref{fig:qae_training_circuit}(b). 

\section{\label{sec:results} Experiments and Results}
In this section we discuss the experiments implementing the classical and quantum data analysis approaches described above for the data compression and classification tasks. 

\subsection{Data compression}

\paragraph*{Classical autoencoder}
The classical neural network autoencoder was implemented with the Keras library of TensorFlow~\cite{tensorflow2015-whitepaper}, and it consists of two dense layers in a 4-2-4 structure as in Fig.~\ref{fig:FIG1}(c), with sigmoid activation function. The input data consists of a time series with 2873893 samples, 25\% of which are used as validation data, and the rest for training. Before training, features were transformed with a MinMax scaler, which scaled each feature to fit in the range $[0,1]$. After the learning phase, the average reconstruction error $\bar{e}$, evaluated as
\begin{equation}
\bar{e} = \frac{1}{M}\sum_{i=1}^M \,\left( \frac{1}{4}\sum_{j=1}^4\frac{|x^{(i,j)}_\text{decoder}-x^{(i,j)}_\text{original}|}{|x^{(i,j)}_\text{original}|} \right)
\label{eq:recontruction_error}
\end{equation}

amounts to 5\%, and in Fig.~\ref{fig:FIG3} we show a comparison of the original against reconstructed data averaged by day, for the validation dataset. As we can see, the decoder shows quite good performance in the reconstruction of the input data for 3 of the 4 variables. For the ‘LIC’ variable, the median of the distribution of the reconstructed data coincides with the one of the original data, though the fluctuations are not very well described. There is no obvious a priori reason for the imperfect reconstruction of this particular variable, and this may well be a shortcoming of the autoencoding approach, which focuses more on the other variables to achieve a good-enough reconstruction scheme. 

In the following step we used the two variables from the compressed layer as input for a supervised classification algorithm, to predict the class assigned at the beginning through the clustering algorithm. We expect that, if the compressed vector is a suitable representation of the input data, a classification algorithm would be able to achieve very good performances.

\paragraph*{Quantum autoencoder}
The quantum autoencoder was simulated using a combination of PennyLane~\cite{Pennylane} with the TensorFlow~\cite{tensorflow2015-whitepaper} interface, as well as Qiskit~\cite{Qiskit}, and the optimization was thus performed using the automatic differentiation techniques implemented by these libraries. While this is only possible when performing a classical simulation of the quantum algorithm, in realistic scenarios of optimizing a quantum circuit on real quantum hardware one can resort to parameter-shift rules~\cite{Schuld2019Gradients, Mitarai2018Learning} to estimate gradients and optimize variational parameters. 

\begin{figure}[htbp]
    \centering
    \includegraphics[width=\columnwidth]{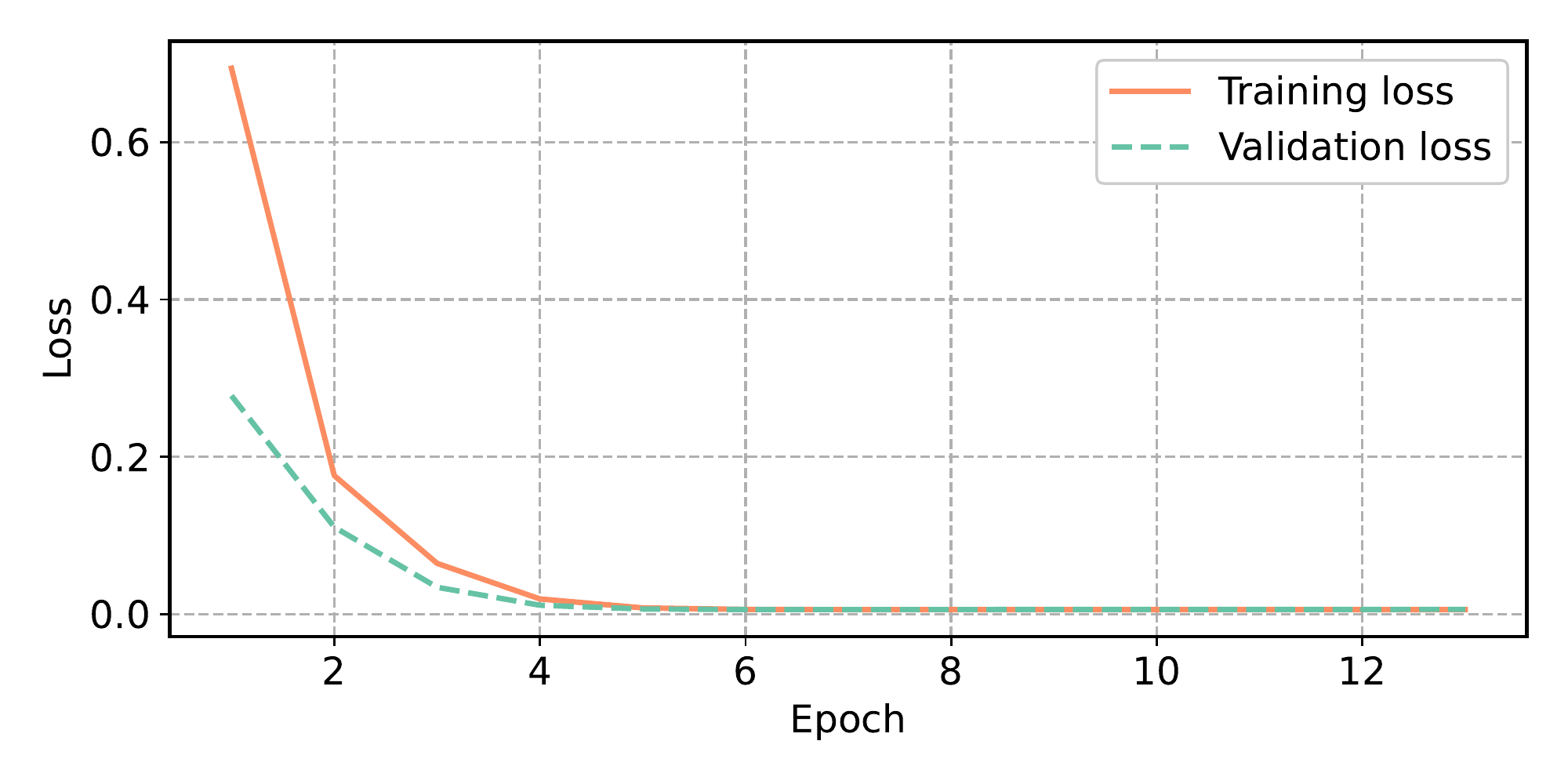}
    \caption{Optimization of the quantum encoder, $\mathcal{E}(\bm{\theta})$, showing the training and validation loss evaluated with data sets containing $10040$ and $520$ samples, respectively. The optimal loss is found at $\mathcal{L}(\bm{\theta}^\ast)=0.0058$.}
    \label{fig:training_loss}
\end{figure}

The variational circuit was trained using the Adam optimizer~\cite{AdamOptimizer} with learning rate set to $0.001$, to update the six variational parameters $\bm{\theta} = (\theta_0, \theta_1, \theta_2, \theta_3, \theta_4, \theta_5)$. The training was performed using mini-batches of size $20$ for a total training set consisting of $10040$ samples. In Fig.~\ref{fig:training_loss} it is shown the optimization process across epochs of learning, both for the training loss, and for a validation set of $520$ samples. Before the phase encoding process, the classical data $\{\vec{x}_i\}_i$ were normalized as $\vec{x}_i\leftarrow \pi \cdot \vec{x}_i/||\vec{x}_i||$. It is clear that the quantum encoder is effectively trained, with the loss reaching the minimum value of $\mathcal{L}(\bm{\theta}^\ast) = 0.0058$.

With a trained encoder, we can now proceed to investigate the quality of the data compression provided by the algorithm. The state of the qubits $A$ and $B$ after the quantum encoder operator consists of a general two-qubit state
\begin{equation}
\begin{split}
\ket{\Psi}_{AB} =\, & a\ket{0}_B\otimes\ket{0}_A + b\ket{0}_B\otimes\ket{1}_A +  \\
& c\ket{1}_B\otimes\ket{0}_A + d\ket{1}_B\otimes\ket{1}_A \, ,
\end{split}
\end{equation}
where, if the encoder has been successfully trained, the probability of measuring qubit $B$ in state $\ket{1}$, $p_1 = |c|^2 + |d|^2$, is much smaller (ideally zero) than the probability of finding it in $\ket{0}$, i.e. $p_1\ll p_0 = |a|^2 + |b|^2$. 
Thus, in order to obtain a compressed pure state for qubit $A$ instead of a mixed one, we could post-select state $\ket{\Psi}_{AB}$ on measuring the trash qubit in state $\ket{0}$. Be $\hat{\Pi}^B_0 = \dyad{0}_B$ the projector on state $\ket{0}$ for system $B$, then the composite state is projected to
\begin{equation}
\begin{split}
    \ket{\Psi}_{AB} \longrightarrow\, & \frac{\hat{\Pi}^B_0 \dyad{\Psi}\hat{\Pi}^B_0}{\Tr_B [\dyad{\Psi}]}\\
    =\, &  \ket{0}_B\otimes \frac{a\ket{0}_A + b\ket{1}_A}{|a|^2+|b|^2} \\
    =\, & \ket{0}_B \otimes \ket{\psi_c}_{A} \, .
\end{split}
\end{equation}

\begin{figure*}
    \centering
    \includegraphics[width=\textwidth]{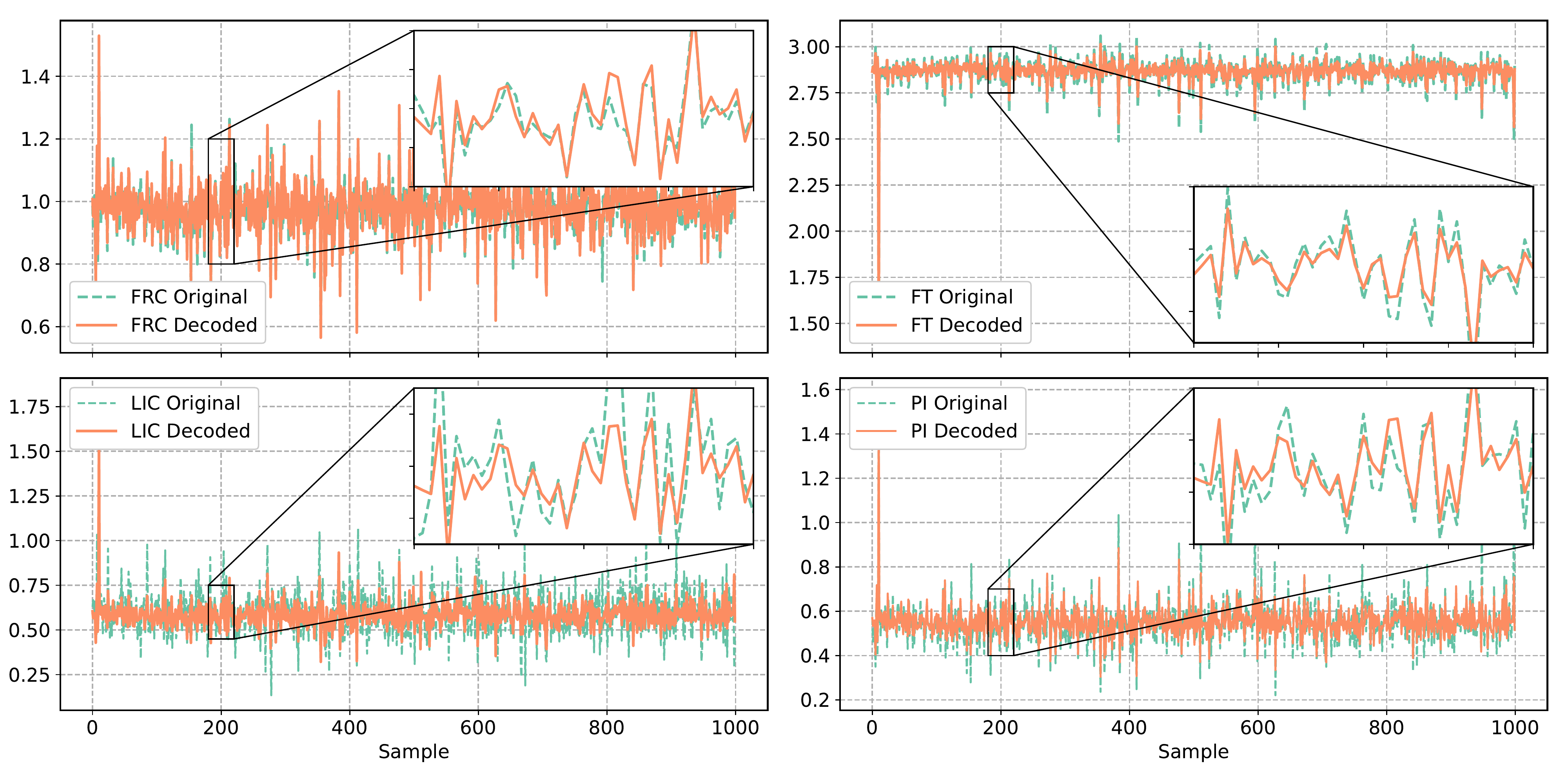}
    \caption{Performances of the quantum autoencoder in a compression and decoding task. Each plot shows one of the input features labeled `FRC', `FT', `LIC', `PI', as reconstructed by the quantum autoencoder (`decoded') confronted with the original sample (`original'). This plots are evaluated on a test set consisting of $M=1000$ samples. The average reconstruction error $\bar{e}$ as defined in the main text~\eqref{eq:recontruction_error}, amounts to $\bar{e}=5.4\%$. This results were obtained using the IBM Qiskit \texttt{statevector\_simulator}.}
    \label{fig:quantum_reconstruction}
\end{figure*}

If we wish to retrieve the original information, now stored in compressed form in the state $\ket{\psi_c}_{A}$ of system $A$ only, we can couple this system to a new qubit initialized in $\ket{0}$, and then apply the quantum decoder, as shown in Fig.~\ref{fig:quantum_autoencoder}. An example of this procedure is shown in Fig.~\ref{fig:quantum_reconstruction}, where the reconstruction performances of the quantum autoencoder are evaluated on a test set consisting of $M=1000$ samples coming from the original dataset. In the case of Fig.~\ref{fig:quantum_reconstruction}, the average reconstruction error (see Eq.~\eqref{eq:recontruction_error}) amounts to $\bar{e} = 5.4\%$, confirming that the quantum autoencoder can successfully compress and retrieve original information with low error.  

However, it is important to stress that these results were obtained using the Qiskit \texttt{statevector\_simulator} from IBM, which allowed us to have direct access to the amplitudes of the quantum states, and thus recover the final phases of the decoded state, $\ket{\varphi_\text{decoder}}=\mathcal{D}(\bm{\theta})(\ket{0} \otimes \ket{\psi_c}_{A})$. In fact, in a real case scenario with a quantum hardware, it is not possible to perfectly retrieve the phases of the decoded state $\ket{\varphi_\text{decoder}}$, since one would need to perform quantum tomography of such state, and even in that case results could only be obtained up to an arbitrary constant, due to quantum measurement outcomes following Born's rule. Thus, while such reconstruction test would prove much harder to be performed on a real device, the results in Fig.~\ref{fig:quantum_reconstruction} obtained with the simulator are still relevant in checking the  inner working of the quantum autoencoder, and check that it is actually able to perform the task it was designed for, even if it is not currently accessible by a real experimenter. 

\begin{figure}[htbp]
\includegraphics[width=\columnwidth]{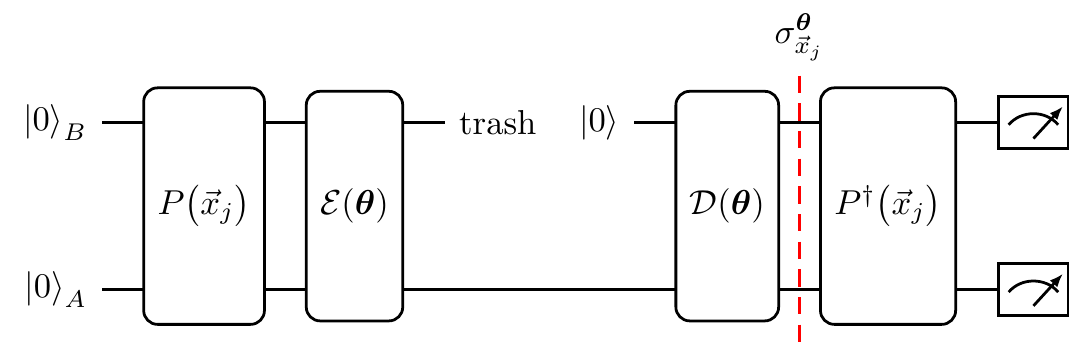}
\caption{Circuit to evaluate the fidelity $F(\rho_{\vec{x}},\sigma^{\bm{\theta}}_{\vec{x}}) = \Tr[\rho_{\vec{x}}\,\sigma^{\bm{\theta}}_{\vec{x}}]$ between the initial pure state $\rho_{\vec{x}} = \dyad{\psi_{\vec{x}}}$ and the generally mixed state $\sigma^{\bm{\theta}}_{\vec{x}}$, obtained through the autoencoding procedure. The fidelity is obtained by counting the number of $\ket{00}$ outcomes. In fact, dropping the subscripts for simplicity, one has $\Tr[P^\dagger \sigma P \dyad{\bm{0}}] = \Tr[\sigma P \dyad{\bm{0}}P^\dagger]= \Tr[\sigma \dyad{\psi}]$.}
\label{fig:FIG8_fidelity}
\end{figure}

There is a second possible approach, which albeit being indirect does not require state tomography and is thus more readily compatible with actual runs on quantum processors. The performances of the quantum autoencoder can be tested measuring the fidelity~\cite{QIT_Wilde} $F(\rho_{\vec{x}},\sigma^{\bm{\theta}}_{\vec{x}})=\Tr[\rho_{\vec{x}}\,\sigma^{\bm{\theta}}_{\vec{x}}]$ between the initial pure state $\rho_{\vec{x}} = \dyad{\psi_{\vec{x}}}$ obtained through phase encoding~\eqref{eq:phase_encoding}, and the generally mixed state obtained through the quantum circuit autoencoder (see Fig.~\ref{fig:quantum_autoencoder})
\begin{equation}
    \sigma^{\bm{\theta}}_{\vec{x}} = {\sf D}(\bm{\theta})\big[\dyad{0}\otimes \Tr_B [{\sf E}(\bm{\theta})[\rho_{\vec{x}}]\big]\, ,
\end{equation} 
where {\sf E}$(\bm{\theta})$ and {\sf D}$(\bm{\theta})$ represents the superoperator corresponding to the encoder $\mathcal{E}(\bm{\theta})$ and decoder $\mathcal{D}(\bm{\theta})$ operators, respectively. Clearly, the larger the fidelity the better, since it corresponds to the quantum autoencoder being able to recreate states that are very close to the initial ones.
Using this figure of merit, post-selecting on the trash subsystem $B$ is not necessary, since qubit $A$ can be directly coupled to a new qubit initialized in $\ket{0}$ to act with the decoder, and then proceed to evaluating $\Tr[\rho_{\vec{x}}\,\sigma^{\bm{\theta}}_{\vec{x}}]$. There are various techniques to evaluate state overlaps on quantum hardware~\cite{Cincio2018Overlap, ManginiQNeuronl2020}, the most common one being the SWAP test, and here we use leverage the so-called \textit{compute-uncompute} method, whose circuit is shown in Fig.~\ref{fig:FIG8_fidelity}. 

Using a test set of $M=1000$ samples, a simulation of the trained quantum autoencoder, even including stochastic measurement outcomes with $10^4$ shots, yields an average fidelity
\begin{equation*}
\mathbb{E}[\Tr[\rho_{\vec{x}}\sigma^{\bm{\theta}}_{\vec{x}}]] = \frac{1}{M}\sum_{j=1}^M \Tr[\rho_{\vec{x}_j}\sigma^{\bm{\theta}}_{\vec{x}_j}] = 0.975\pm 0.001\,,
\end{equation*}
which confirms again that the proposed variational quantum autoencoder is able to compress and later decode information. 

\subsection{\label{sec:classification} Classification}

\begin{figure}
    \centering
    \includegraphics[width=\columnwidth]{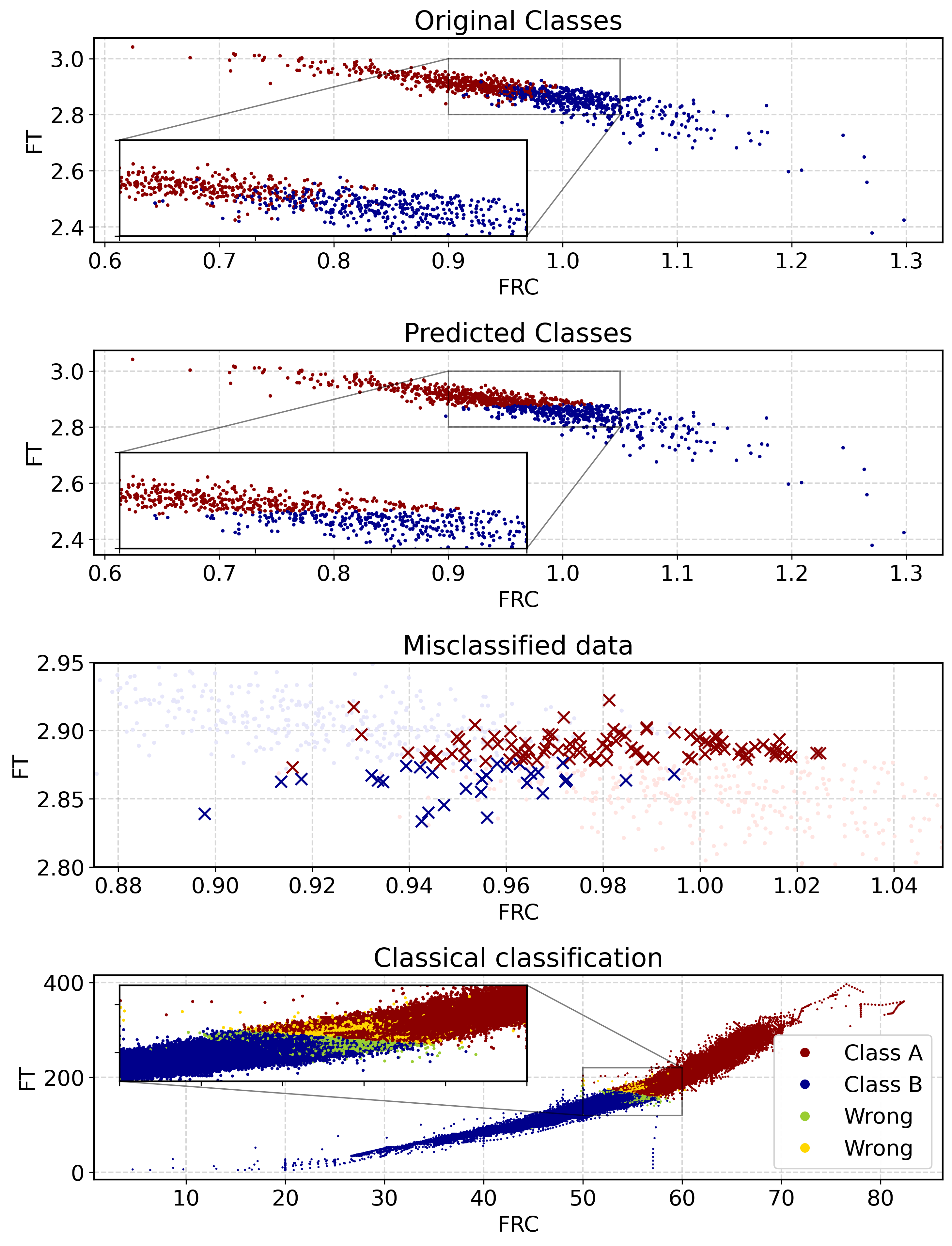}
    \caption{Results of the classification task performed by the quantum classifier, for a test set of size $10^3$ samples. \textbf{(a)} Plot of the original data with the color indicating the two different classes. Note that for simplicity, only the FRC and FT features are shown. \textbf{(b)} Label assigned by the trained quantum classifier. \textbf{(c)} Focus on the data that are mislabeled by the classifier. The color indicates the label assigned by the quantum classifier, and the ``cross" marker means that the data were misclassified. Note that these samples lay on the border of separating the two classes. The accuracy, evaluated as the percentage of correctly classified data, amounts to $87.4\%$. \textbf{(d)} Result of the classification using the classical autoencoder followed by a KNN clustering procedure. Note that the axis are different from the quantum case due to normalization of the features. In this case the classification accuracy amounts to $89.7\%$}
    \label{fig:quantum_classification}
\end{figure}

\paragraph*{Classical classifier}
The supervised classification algorithm used is the KNeighborsClassifier as implemented in \texttt{scikit-learn}. KNeighborsClassifier assigns the class to a point from a simple majority vote based on the $k$ nearest neighbors of that point. The number of nearest neighbors is a parameter of the algorithm, and after some trials we fixed it at $k=100$, which correspond to an optimal trade-off between performances and computational efficiency. The lowest panel of Fig.~\ref{fig:quantum_classification} shows the results of the classification, which is now anticipated but discussed later in comparison with the quantum algorithm results. In red and blue are the points correctly classified, while in yellow and green are those which were misclassified. 
The classification accuracy, evaluated as the percentage of correctly classified data, reach a remarkably high value of $89.7\%$, indicating that the compressed vector is able to summarize the information carried by the input data. 

\paragraph*{Single qubit quantum classifier}
Once the quantum autoencoder has been trained to learn a compressed representation of the original information, the compressed quantum state can be used as input for a classification task. We expect that, if the compressed information is a suitable representation of the input data, the classification algorithm would be able to learn the classes assigned to the full-size input data through the clustering algorithm described in Sec.~\ref{sec:case_study}. 
To do so, we can use the information-carrying qubit obtained with the encoder $\mathcal{E}(\bm{\theta})$, as input to a quantum classifier which is trained to learn the desired clustering of the original data. A quantum classifier is made of two parts: a trainable parametrized operation, $U$, which tries to map inputs belonging to different classes in two distant regions of the Hilbert space; and a final measurement, which is used to extract and assign the label. 
Since we are dealing with a single qubit classifier, the most general transformation on a qubit is represented by the unitary matrix
\begin{equation}
    U\left(\alpha,\beta,\gamma\right)=
    \begin{bmatrix}
    \cos(\alpha/2) & e^{-i\gamma}\sin(\alpha/2)\\
    e^{i\beta}\sin(\alpha/2) & e^{i(\beta+\gamma)}\cos(\alpha/2) \, .
    \end{bmatrix}
\label{eq:u3}
\end{equation}
Thus, it is reasonable to use such operation as the trainable block of the classifier, since it ensures the greatest flexibility. Actually, as discussed later, the angle $\beta$ in Eq.~\eqref{eq:u3} does not influence the measurement statistics of the qubit, hence it has no influence on the training of the classifier. For this reason, it is kept fixed at $\beta=0$, and the actual trainable gate used is $U(\alpha,0,\gamma)=U(\alpha, \gamma)$. 

As for the label assignment, since the measurement process of a qubit has only two possible outcomes, these are interpreted to be the two possible value for the labels, that is ``Class A" and ``Class B" described in Sec.~\ref{sec:case_study}. 
More in detail, a label is assigned based on a majority vote on multiple shots of the same quantum circuit: an input is assigned to “Class A” if the majority of measurement gave $\ket{0}$ as outcome, “Class B” otherwise. Formally, be $\rho^A_{\vec{x}} = \Tr_B [\mathcal{E}(\bm{\theta})(\dyad{\psi_{\vec{x}}})\mathcal{E}(\bm{\theta})^\dagger]$, the compressed quantum qubit, then the label is assigned following the decision rule:
\begin{equation}
\hat{y}_i=
\begin{cases}
0\quad \text{if}\quad p_0=\Tr[\dyad{0}U\rho^A_{\vec{x}_i} U^\dagger]\geq0.5\\
1 \quad \rm{otherwise}
\end{cases}
\end{equation}
where $p_0$ denotes the probability that the measurement yields $\ket{0}$ outcome. As mentioned earlier, one can check easily that $p_0$ does not depend on the angle $\beta$ of the unitary $U(\alpha, \beta, \gamma)$, and for this reason it is set to zero, yielding the variational unitary $U(\alpha,0,\gamma)=U(\alpha, \gamma)$.  

The loss function used to drive the training of the unitary $U(\alpha, \gamma)$ is the categorical cross entropy, defined as
\begin{equation}
\mathcal{L}(y_i,\hat{y}_i)=-(1-y_i)\log(1-\hat{y}_i)\ -\ y_i\log(\hat{y}_i) \, ,
\end{equation}
where $y_i$ is the correct label, and $\hat{y}_i$ is the label assigned by the quantum classifier, and the optimizer used is COBYLA~\cite{COBYLAPowell1994} as implemented in SciPy's python package~\cite{2020SciPy-NMeth}.

Figure~\ref{fig:quantum_classification} shows the results of the classification obtained after the optimization of the variational parameters $(\alpha, \gamma)$, for a test set of $M=10^3$ samples. The accuracy, measured as the ratio of correctly classified to total samples, is measured to be 87.4\% when evaluated with exact simulation of the quantum circuit. As clear from the figure, the misclassified data are only those located near the edge connecting the two classes. In fact, in this region, the samples are not neatly divided but rather a blurred border exists. On the contrary, the quantum classifier, given its relatively simple structure, learns essentially a straight cut of the data in this region, thus committing some labeling errors. This should not come as a surprise however, since it is known from the literature that, if not using expressivity enhancement techniques like \textit{data re-uploading}, a single qubit classifier can only learn simple functions (i.e. sine functions) of the input data~\cite{Schuld2020Encoding, Perez2020Reuploading, PerezSalinas2020datareuploading, LaRose2020Robust, Gratsea2021exploring, Lloyd2020Embeddings}. In addition, remind that the classical data are loaded onto the quantum states by means of rotations, hence the dependence of the classification on the original classical data is strictly non-linear, with the presence of the encoder scrambling information even more.
It is interesting to notice that the classification performances remain stable even when including sources of noise, such as stochastic measurement outcomes. In this case, using $n_{shots}=1024$, the accuracy amounts to about $82.5\%$, with the uncertainty due to the stochastic nature of the simulation. In addition, the classifier proves robust even with tests performed on real quantum hardware. In fact, the circuit for the trained classifier was tested against IBM's \texttt{ibmq\_x2} quantum chip (accessed May 2021) but with a smaller test set of $75$ samples, due to limitations in the device usage. In this case, using $1024$ shots per circuit, and averaging on 5 executions with different test samples, the classification accuracy (evaluated again as the percentage of correctly classified data) was found to be $(82.3\pm1.3)\%$, indeed very close to the simulation including only measurement noise, and not much different from the noiseless result. 

\section{\label{sec:conclusion} Summary and outlook}
We have presented a direct comparison between quantum and classical implementations of a neural network autoencoder, followed by a classifier algorithm, applied to sample real data coming from one of Eni's plants, in particular from a first stage separator. While the achievement of a clear quantum machine learning advantage with variational algorithms is still disputed~\cite{HuangQMLAdvantage2021,Huang2020Power,Abbas2020Power}, this work sets a milestone in the field of quantum machine learning, since it is one of the first examples of direct application of quantum computing software and hardware to analyse real data sets from industrial sources. 

As a first step, we have implemented and analyzed the performance of a variational quantum autoencoder to compress and subsequently recover the input data. We verified its performances using full simulation of the wavefunction, which allowed us to evaluate the average reconstruction error to about $\bar{e}=5\%$ ---essentially identical to the classical autoencoder--- thus confirming the capability of the quantum autoencoder to effectively store a compressed version of the original data set, and then being able to recover it. In addition, we also checked the correctness of the quantum autoencoding procedure by evaluating the quantum fidelity between original and decoded quantum states, which were again found to be very similar to each other even in the presence of simulated stochastic measurement noise. Once the optimal parameters for the quantum autoencoder were determined during the training phase, we used the compressed quantum state as input to a quantum classifier, with the goal of  performing a binary classification task. The algorithm achieved an accuracy above $87\%$, absolutely comparable to that achieved in the classical setting using the neural network autoencoder followed by a nearest-neighbors classifier, thus indicating again that the quantum algorithm is able to correctly compress the relevant information of the input data. We also tested the performance of the full quantum pipeline (given by the quantum autoencoder plus the classifier) on actual and currently available IBM quantum hardware, obtaining a classification accuracy of $82\%$, which is only slightly smaller than the ideal result. 

The small size of current quantum devices and their relatively high noise levels make it hard to run actually relevant and large scale computations, thus making an effective quantum advantage out of reach. We provided, on the other hand, a successful proof-of-concept demonstration that an original quantum autoencoder and a quantum classifier can actually reach the same level of accuracy as standard classical algorithms, on a data set that is sufficiently low dimensional to be handled on actual near-term quantum devices. In addition, it is worth emphasizing that the quantum autoencoder allows to obtain results that are quantitatively comparable to the classical algorithm by using only 6 parameters instead of 16, thus displaying an increased efficiency in terms of number of trainable parameters already reached on NISQ devices. With continuing progress in quantum technologies and quantum information platforms, we envision the execution of the very same quantum algorithms on larger scales, possibly reaching the threshold for a classically intractable problem. We believe these results take the first foundational steps towards the application of usable quantum algorithms on NISQ devices for industrial data. 

\begin{acknowledgments}
We acknowledge use of the IBM Quantum Experience for this work. The views expressed are those of the authors and do not reflect the official policy or position of IBM company or the IBM-Q team. We wish to thank Eni's management, who gave the permission to publish this paper. This work was supported by Eni S.p.A. through the research program ``Research, Development and Analysis Activities supporting the Innovation'' under Contract No. 2500034935, and by 
the Italian Ministry of Education, University and Research (MIUR) through the ``Dipartimenti di Eccellenza Program (2018-2022)''
\end{acknowledgments}

\section*{Disclosure of potential conflicts of interest}
The authors declare no competing interests.

\bibliographystyle{apsrev4-2}
\section*{References}
\bibliography{bibliography}

\end{document}